\documentclass[aps,prb,twocolumn,groupedaddress]{revtex4}
\usepackage{graphicx}% Include figure files

\begin{document}

\newcommand{\be}{\begin{equation}}
\newcommand{\ee}{\end{equation}}
\newcommand{\bea}{\begin{eqnarray}}
\newcommand{\eea}{\end{eqnarray}}
\newcommand{\nt}{\narrowtext}
\newcommand{\wt}{\widetext}

\title{Excitonic pairing between nodal fermions}
\author{D. V. Khveshchenko and W. F. Shively}
%\email{shively@physics.unc.edu}
\affiliation{Department of Physics and Astronomy, University of North Carolina,
Chapel Hill, North Carolina 27599, USA}

\begin{abstract}
We study excitonic pairing in nodal fermion systems characterized
by a vanishing quasiparticle density of states at the pointlike
Fermi surface and a concomitant lack of screening for long-range interactions.
By solving the gap equation for the excitonic order parameter, we
obtain a critical value of the interaction strength for
a variety of power-law interactions and densities of states.
We compute the free energy and analyze possible phase transitions,
thus shedding further light on the unusual pairing properties of this peculiar
class of strongly correlated systems.
\end{abstract}

%*** insert suggested PACS numbers in braces on next line?

%Note!  Now \maketitle must follow abstract

\maketitle

%this executes the '\maketitle' macro

%\begin{multicols}{2}
%should also omit out the 'multicols'{2} here as well - see section 5.8

Since its foundation some fifty years ago, Fermi-liquid theory
has been providing a convenient framework for the analysis of conventional metals.
Nowadays, however, the quest for departures from classic Fermi-liquid behavior
has become one of the central themes in modern condensed matter theory.

In a generic one-dimensional (1D) many-fermion system,
an arbitrarily weak short-range repulsive interaction is known to
completely destroy the Fermi liquid, thereby giving rise to
so-called Luttinger behavior. In higher dimensions (2D and 3D),
the Fermi liquid is believed to be more robust, although it is
not expected to remain absolutely stable. While in the case of
short-range repulsive interactions any deviations from the Fermi liquid
are likely to be limited to the regime of infinitely strong couplings,
the long-ranged interactions are
considered to be generally capable of destroying the Fermi liquid even at finite couplings.

In that regard, a particularly promising case appears to be the class of systems which possess
isolated (``nodal'') Fermi points where the quasiparticle density of states
(DOS) vanishes. As a result, any long-ranged interaction
(whether the bare Coulomb one or that mediated by soft
collective modes proliferating near an emergent instability) remains largely unscreened.

Examples of the nodal fermion systems are provided by
degenerate semimetals, such as graphite or bismuth, where the
conduction and valence bands touch each other at isolated points in
the Brillouin zone. In a monolayer of graphite, for example, the two-dimensional
fermion excitations with the momenta close to the nodal points
have a linear dispersion, and, therefore, can be described in terms of
a Dirac-like effective theory. \cite{semenoff}

Amongst other potential candidates to the nodal liquid scenario are
$p$ (such as He$_3$) and $d$ (such as cuprates)
superfluids where the quasiparticle gap has isolated nodes.
Yet another example is provided by charge density wave (CDW) insulators,
such as dichalcogenides, which are thought to have $f$-wave
CDW order. \cite{antonio}

On general grounds, one might expect behavior manifested by systems
supporting nodal quasiparticles to be markedly
different from that of fermions with an extended Fermi surface
from which a bare long-range interaction would undergo strong screening.
In particular, it is conceivable that pairing instabilities
which, depending on the sign of the interaction,
occur either in the particle-particle (Cooper) or particle-hole (Peierls) channels,
develop differently from the conventional BCS scenario.

Barring a handful of counter-examples, most of the previous
studies of fermion pairing were limited to the simplest
BCS case where the momentum dependence of the
corresponding order parameter (except, possibly,
for its angular dependence) would be routinely neglected.
However, while being generally justified in the case of normal metals
with their intrinsic screening, the BCS solution
may no longer be applicable in the nodal liquids.

In the present paper, we address the general problem
of pairing between the nodal fermions with a generic
dispersion $\xi_p\propto p^\eta$. The latter corresponds to
the power-law DOS $\nu(\epsilon)\propto\epsilon^\beta$
which vanishes at the nodal Fermi point located at ${\bf p}=0$
and is governed by the exponent $\beta={D/\eta-1}$.

Although below we focus on excitonic pairing pertinent
to the case of repulsive interactions, much of the following discussion
can be readily adapted to the case of the Cooper instability
caused by attraction.

In the particle-hole representation, the inverse fermion Green function reads
\begin{equation}
\hat{G}^{-1}(\omega,{\bf p})=\omega-\mu-{\hat \tau}_3\xi_p+{\hat \tau}_1\Delta_p
\end{equation}
where ${\hat \tau}_i$ is the triplet of the Pauli matrices acting in the particle-hole
space, $\mu$ is the chemical potential, and $\Delta_p$ is the gap function
which describes the anticipated spatially uniform 
(see the conclusions for a discussion of this assumption) 
pairing in the Peierls channel caused by a long-ranged repulsive potential with the Fourier
transform $U^{(0)}({\bf q})=g_0/q^{\eta\alpha}$.

When the nodal fermion polarization $\chi(\omega, {\bf q})$ is taken into
account, the renormalized effective interaction assumes the form
\be
U(\omega, {\bf q})=\frac{g_0}{{q}^{\eta\alpha}+g\chi(\omega,{\bf q})}.
\ee
At low (compared to a characteristic scale 
$\Lambda$ which is of order the maximum span of the Brillouin zone)
momenta and still lower frequencies ($\omega/q^\eta\to 0$ and
$q\to 0$) the one-loop polarization behaves as $\chi(0,{\bf q})\propto q^{D-\eta}$.
Thus, under the condition $\alpha\leq\beta$, the effective interaction (2)
retains its bare form (except for a finite renormalization of the coupling
constant $g_0\to g_0\left[1+g_0\chi(0,{\bf q})/q^{\eta\alpha}\right]^{-1}|_{q\to 0}$ and an additional frequency dependence for
$\alpha=\beta$). By contrast, for $\alpha>\beta$ the overscreened effective interaction (2) acquires
a universal form governed by the fermion polarization, thereby giving
rise to the behavior which is qualitatively similar to that occurring at $\alpha=\beta$.

The gap equation is given by the ${\hat \tau_1}$-component of the
Dyson-Schwinger equation for the fermion propagator (1)
\bea
\hat{G}^{-1}(\omega, {\bf p})=\hat{G}^{-1}_0(\omega, {\bf p})+\nonumber\\
i\int\frac{d\Omega}{(2\pi)}\sum_{\bf q}
U(\Omega-\omega, {\bf p}-{\bf q})\hat{G}(\Omega, {\bf q})
\eea
where ${\hat G}_0(\omega, {\bf p})$ is the bare propagator with $\Delta_p=0$.

In what follows, we focus on the regime $\alpha\leq\beta$ where,
conceivably, the polarization operator may still become important
at the momenta of order the upper cutoff set by $\Lambda$. We find, however,
that a rapid decrease of $\Delta_p$ at large momenta (see below) makes 
the gap function robust against the inclusion of such high-energy polarization effects.
By the same token, a polarization-related frequency dependence (retardation) of the effective interaction 
does not have a significant impact on the solutions of Eq.(1) and, therefore, will be neglected from now on. 

Taking into account the above arguments,
we perform the frequency integration in Eq.(3) and take its trace
with ${\hat \tau}_1$, thereby arriving at the integral equation
for the momentum (albeit not the frequency) dependent
gap function
\be
\Delta_{\bf p}=\sum_{{\bf q},\mp}U(0,{\bf p}-{\bf q}){(\mp)\Delta_{\bf q}\over 2E_q}\theta({\mu\mp E_{\bf q}}),
\ee
where $E_{\bf q}={\sqrt{{\xi_{\bf q}}^2+\Delta_{\bf q}^2}}$ is the quasiparticle energy, and
the sum is taken over all the occupied states in both the valence
($\Omega=-E_{\bf q}$) and conduction ($\Omega=E_{\bf q}$) bands.

Notably, for any $\alpha>0$, a nontrivial momentum dependence of the
integral kernel in Eq.(4)
rules out the possibility of the standard BCS solution $\Delta_{\bf p}=$ const.
While resisting an exact analytical solution for general values
of $\alpha$ and $\beta$, Eq.(4) can still be amenable to numerical analysis.
However, we find it more physically transparent to resort to an approximate
analytical approach which generalizes that of Refs. 3 and 4 and allows one to convert an integral equation into a simpler linear differential one.
This method elaborates
on the earlier analyses \cite{appelquist} of the phenomenon of chiral
symmetry breaking in 3D relativistic fermion systems that can be thought of as a
Lorentz-invariant analog of excitonic pairing.

We start off by observing that, at large momenta, Eq.(4) can be simplified by replacing
the kernel $U({\bf p-q})$ with either $U({\bf p})$ or $U({\bf q})$,
depending on the relative magnitude of the momenta $\bf q$ and $\bf p$. Next, by changing
the integration variable from the momentum $\bf p$ to the quasiparticle energy
$\epsilon=E_{\bf p}$ and repeatedly
differentiating both sides of Eq.(4), we obtain the linear differential equation
\be
{d^2\Delta(\epsilon)\over d\epsilon^2}+{\alpha+1\over
\epsilon}{d\Delta(\epsilon)\over d\epsilon}+g{\alpha\Delta(\epsilon)\over
\epsilon^{2+\alpha-\beta}}=0,
\ee
where $g=g_0/[2^{D-1}\pi^{D/2}\Gamma(D/2)]$.

In order for Eq.(5) to be consistent with the original integral Eq.(4),
it has be complemented with the boundary conditions
\be
\left.{d\Delta(\epsilon)\over d\epsilon}\right|_{\epsilon=\max[\Delta,\mu]}=0
\ee
and
\be
\left.\Delta(\epsilon)+{\epsilon\over \alpha}{d\Delta(\epsilon)\over dE}\right|_{\epsilon=\Lambda}=0.
\ee
In what follows, we refer to Eqs.(6) and (7), respectively, as the
infrared (IR) and ultraviolet (UV) boundary conditions.

Although the above approximations can be fully justified only at
large momenta, a comparison between the numerical analyses of the original
gap equation (4)
and the relatively simple approximate solutions of the linearized equation (5) (see below)
shows that the latter provide a faithful representation of the solutions
to Eq.(4) down to the lowest energies $\epsilon\sim\Delta$.

We first consider the case of half filling ($\mu=0$). The linearized Eq.(5)
can be interpreted as the radial Schr\"{o}dinger equation of the zero energy state in the
fictitious $(\alpha+2)$-dimensional spherically symmetrical 
potential $V(R)=g\alpha/R^{2+\alpha-\beta}$. This
analogy suggests that for $0\leq\beta-\alpha<1$ the zero energy
solution occurs for couplings $g$ in excess of a finite threshold value $g_c$.

This prediction is supported by a direct analysis of Eq.(5)
which, for $\alpha>\beta/3$,
appears to have an approximate solution of the form	
\be
\Delta\left(\epsilon\right)=
\Delta\left(\frac{\Delta}{\epsilon}\right)^{\gamma(\epsilon)}
{\sin\left[\Phi(\epsilon)+\delta\right]\over \sin\delta}
\ee
where the prefactor is chosen to satisfy the natural normalization condition
$\Delta(\epsilon)|_{\epsilon=\Delta}=\Delta$, and
the phase of the trigonometric function is given by the expression
\be
\Phi(\epsilon)={1\over r}{\sqrt {\alpha g\epsilon^{2r}-\kappa(\epsilon)/4}}
\left[1-(\Delta/\epsilon)^r\right]
\ee
Here $r=(\beta-\alpha)/2$, the phase shift
$\delta=\tan^{-1}[{{\sqrt {4\alpha g\Delta^{2r}-\kappa(\Delta)}}/\alpha}]$
for $\Delta>\left[\kappa(\Delta)/4\alpha g\right]^{1/2r}$, otherwise $\delta=0$,
is determined by the IR boundary condition,
and the slowly varying functions $\gamma(\epsilon)$ and $\kappa(\epsilon)$
attain the following asymptotic values:
\be
\gamma(\Delta)=\alpha/2, ~~~~~~\gamma(\Lambda)=(\alpha+\beta)/4
\ee
and
\be
\kappa(\Delta)=
\alpha^2-(\alpha-\beta)^2, ~~~~~~\kappa(\Lambda)=\alpha^2-{1\over 4}(\alpha-\beta)^2.
\ee
The maximum (zero-momentum) value of the gap function $\Delta=\Delta(0)$
can be determined from the UV boundary condition (7), which now assumes the form
\be
\tan[\Phi(\Lambda)]+{4\over 3\alpha-\beta}
{\alpha{\tilde g}-\kappa(\Lambda)(\Delta/\Lambda)^r/4\over
{\sqrt {\alpha{\tilde g}-\kappa(\Lambda)/4}}}=0 ,
\ee
where ${\tilde g}=g\Lambda^{2r}$.

For $0<\alpha<\beta$, Eq.(12) yields
\be
\Delta\sim \Lambda ({\tilde g}-{\tilde g}_{c})^{1/r}
\ee
for couplings greater than the critical value
\be
{\tilde g}_{c}\approx{1\over 16 \alpha}\left[(\alpha+\beta)^2+{\pi^2(\beta-\alpha)^2}\right]
\ee
In the marginal case $\alpha=\beta>0$ and for $g>g_{c}=\alpha/4$
the solution (8) takes the form
\be
\Delta(\epsilon)=\Delta\left(\frac{\Delta}{\epsilon}\right)^{\alpha/2}
{\sin\left[{\sqrt {\alpha g-\alpha^2/4}}\ln(\epsilon/\Delta)+\delta\right]\over \sin\delta},
\ee
where $\delta=\tan^{-1}{\sqrt {4(g/\alpha)-1}}$. 

In the case $\alpha=\beta=1$,
Eq.(15) reproduces the solution that was obtained in Refs. 3 and 4.
In this case, the gap given by Eq.(12) shows a behavior reminiscent of that
at the Kosterlitz-Thouless (KT) transition in $XY$-symmetrical 2D systems
\be
\Delta=\Lambda \exp\left(-{2\pi-4\delta\over {\sqrt {4\alpha(g-g_c)}}}\right).
\ee
A similar behavior has been previously found in the context of 3D relativistic
chiral symmetry breaking, in which it was conjectured to signal
the onset of a conformally invariant critical regime.\cite{appelquist} 
The present discussion shows that the KT-like behavior (16) is not specific
to Lorentz-invariant systems. Rather, it appears to be characteristic
of the scale invariance of the underlying ``radial'' gap equation (5).

Next, we introduce a finite chemical potential $\mu$, which, depending on the sign,
represents a finite density of either particles or holes. The occupation
factor $\theta(\mu\mp E_q)$ in the right-hand side of Eq.(4)
sets a lower limit of momentum integration.

A straightforward analysis shows that at $\epsilon\lesssim\mu$ the
solution $\Delta(\epsilon)$ of the gap equation levels off and becomes 
independent of energy. For $\Delta<\mu$ we then impose the normalization
condition $\Delta(\epsilon)|_{\epsilon=\mu}=\Delta$, thereby
arriving at the approximate formula for
the counterpart of Eq.(8)
\be
\Delta(\epsilon)=\Delta\left(\frac{\mu}{\epsilon}\right)^{\gamma(\epsilon)}
{\sin\left[\Phi(\epsilon)+\delta\right]\over \sin\left[\Phi(\mu)+\delta\right]},
\ee
whereas for $\Delta>\mu$ the solution (8) remains essentially intact.
In particular, the marginal (scale-invariant) case $\alpha=\beta$ is now described by the gap function
\be
\Delta(\epsilon)=\Delta\left(\frac{\mu}{\epsilon}\right)^{\alpha/2}
{\sin({\sqrt {\alpha g-\alpha^2/4}}\ln{\epsilon/\Delta}+\delta)\over
\sin({\sqrt {\alpha g-\alpha^2/4}}\ln{\mu/\Delta}+\delta)}.
\ee
After having found the approximate solutions to the gap equation,
we now proceed to elucidate the nature of the quantum phase transition
which results in the opening of the gap. To that end, we use Eqs.(8) and (18)
to compute the free energy functional $F(\Delta,\mu)$.

The presence of the long-ranged forces makes it
rather challenging to utilize the conventional Luttinger-Ward approach
(for its recent application to a very detailed analysis of
the non-BCS energy-dependent superconducting pairing, see Ref. 6).
Specifically, a straightforward implementation of
this method gives rise to a strongly divergent double momentum integral
$\sum_{{\bf p},{\bf q}}\Delta_pU^{-1}(0,{\bf p}-{\bf q})\Delta_q$
whose kernel is given by the (generally, highly nonlocal) inverse
operator $U^{-1}(0, {\bf p})$.

Moreover, an unequivocal identification of the solutions to the gap equation as minima
(as opposed to maxima) of $F(\Delta,\mu)$ requires one to compute
this functional for all $\Delta$, whereas the previous applications
of the Luttinger-Ward and similar techniques have been limited to
the evaluation of the free energy $F(\Delta_0,\mu)$ at its absolute minimum $\Delta_0$
(``condensation energy'').\cite{chubukov2}

As an alternate approach, we employ a field-theoretical technique
which has been previously used to investigate the conjectured excitonic instability
in pyrolytic graphite.\cite{kiev} This method makes use of the source function
 (see Ref. 4 and references therein)
\be
J(\Delta)=\left.\left(\Delta(\epsilon)+{\epsilon\over \alpha}{d\Delta(\epsilon)\over d\epsilon}\right)\right|_{\epsilon=\Lambda}
\ee
and the order parameter
\be
\sigma(\Delta)=Tr[{\hat \tau_1}\int {d\omega\over 2\pi}\sum_{\bf q}{\hat G}
(\omega,{\bf q})]=\nonumber\\
\left.{\epsilon^{\alpha}\over g_0}\Delta(\epsilon)\right|_{\epsilon=\Lambda}
\ee
which both determine the sought-after (regularized) effective potential
given by the formula
\be
F(\Delta,\mu)=
\int^\Delta_0d\Delta^\prime{d\sigma(\Delta^\prime)\over d\Delta^\prime}
J(\Delta^\prime)+F(0,\mu).
\ee
The integration constant $F(0,\mu)$ is to be found
from the equation for the density of excess particles
\be
-{\partial F(\Delta,\mu)\over \partial \mu}=n(\mu)=\sum_{{\bf p},\mp}
\theta(\mu\mp E_p).
\ee
In order to illustrate this approach on a well-known example,
we first consider the standard BCS pairing between
fermions with a short-range interaction
potential and a finite density of states ($\alpha=\beta=0$).

For $\mu=0$ and an arbitrarily weak repulsion,
the gap equation (4) yields the conventional BCS-type solution
$\Delta(\epsilon)=\Delta_0=\Lambda e^{-1/g}$.

For $\mu<\Delta$ the free energy reads
\be
F_>(\Delta, \mu)={g\Delta^2\over 4g_0}
\left({1\over g}-{1\over 2}-\ln{\Lambda\over \Delta}\right),
\ee
whereas for $\Delta<\mu$ one obtains
\bea
F_<(\Delta, \mu)={g\Delta^2\over 4g_0}\left({1\over g}-{1\over 2}-
\ln{\Lambda\over \mu+{\sqrt {\mu^2-\Delta^2}}}\right)\nonumber\\
-{g\over 2g_0}\mu{\sqrt {\mu^2-\Delta^2}},
\eea
the last term being proportional to the particle density
$n={1\over 2}{\sqrt {\mu^2-\Delta^2}}$.

The free energy given by Eqs.(23,24) and plotted in Fig.1
exhibits a minimum at $\Delta=\Delta_0$ for all $\mu<\mu_{c}=\Delta_0/{\sqrt 2}$
and $g<2$. Upon increasing $\mu$ past the critical value $\mu_{c}$, the minimum
jumps discontinuously to the trivial solution $\Delta=0$,
consistent with the anticipated first order nature 
of the transition in question, which is known to be equivalent to the 
superconducting BCS transition in an external magnetic field.

In the case of Fermi liquids with
screened repulsive interactions, the same scenario describes the excitonic instability. 
It should be noted, however, that the transition driven by a
variable density (which provides a more appropriate picture
of the system, in which the number of carriers is not fixed)
can still be of second order.\cite{volkov}

This given example demonstrates that by merely evaluating the
free energy at the extremal points of $F(\Delta,\mu)$, one may not always
be able to distinguish
the true absolute minimum from a local one
or, worse yet, from a maximum. In that regard, it is worth mentioning the apparent confusion
over the nature of the so-called ``nontrivial'' solution $\Delta={\sqrt {2\Delta_0\mu-\Delta^2_0}}$
that was claimed in Ref. 7 to exist for $\mu>\Delta_0/2$.
This solution was also invoked in some of the recent studies of the conjectured (weakly 
ferromagnetic) excitonic insulating state in hexaborides.\cite{rice}

However, it can be immediately seen from Fig.1 that the extremal point in question
is, in fact, a local maximum of the BCS free energy functional. Therefore,
it could not have provided a viable solution to the 
problem of excitonic pairing in a doped semimetal
even if the hexaborides belonged to this class of materials
(which is now considered rather unlikely from the experimental point of view \cite{bennett}).

As the next example, we discuss the case of a short-range
potential and a linear density of states ($\alpha=0$, $\beta=1$).
Similar to the BCS case, the constancy of the gap function
makes the solution to the gap equation [$\Delta(\epsilon)=\Delta_0=g^{-1}_{c}-g^{-1}$
for $\mu<\Delta$] trivial.

For $\mu<\Delta$ we obtain
\be
F_>(\Delta, \mu)= {g\over 2g_0}{\Delta^2}\left({1\over 3}
\Delta-{1\over 2}\Delta_0\right)
\ee
while for $\Delta<\mu$ the free energy incorporating an excess
electron density $n(\mu)={1\over 2}(\mu^2-\Delta^2)$ takes the form
\be
F_<(\Delta, \mu)={g\over 2g_0}\left(-{1\over 6}\mu^3+{1\over 2}\mu\Delta^2-{1\over 2}\Delta_0\Delta^2\right).
\ee
Notably, the terms of order
$\Delta^2\Lambda$ which are present in both the
kinetic and potential contributions cancel each other, and
the condensation energy is governed by the fermion states with energies $\epsilon\sim\Delta$.

As follows from Eqs.(25,26), the minimum of the free energy plotted in Fig.2 exists for all
$g>g_{c}=1/\Lambda$. Interestingly, at $\mu_c=\Delta_0$ 
the mean-field free energy becomes independent of $\Delta$ in the entire range
$0\leq\Delta\leq\Delta_0$, thus signaling the possibility of
a continuous transition. It is, however, conceivable that upon accounting for the fluctuations about the 
mean-field solution one would find the transition in question to be of
(possibly, weakly) first order. 

Being primarily concerned with the properties
of the systems governed by long-ranged interactions,
we now move to the case of a linear density of states and the unscreened
Coulomb interactions ($\alpha=\beta=1$). Unlike in the previous examples, we now find a
nontrivial gap function that conforms to the ``scaling-invariant''
momentum dependence (15). Computing the free energy for $\Delta<\mu$ we obtain
\bea
F_<(\Delta,\mu)={g\over 2g_0}[I_<(\Delta,\mu)-I_<(\Delta_0,\mu)\nonumber\\
-{1\over 2}\int^\mu_\Delta d\omega(\omega^2-\Delta^2)],
\eea
where 
\be
I_<(\Delta,\mu)=\mu\int^\mu_0d\omega\omega{\ln{(\Delta_0/\omega)}\left[5-\ln{(\Delta_0/\omega)}\right]\over
(2+\ln{\mu/\omega})^2}
\ee
whereas for $\Delta>\mu$ the free energy reads
\bea
F_>(\Delta,\mu)\nonumber\\
={g\over 2g_0}[I_>(\Delta,\mu)-I_>(\mu,\mu)+I_<(\mu,\mu)-I_<(\Delta_0,\mu)],
\eea
where 
\be
I_>(\Delta,\mu)=-{2\over 3}\Delta^3-2\Delta^3\ln{\Delta_{0}\over\Delta}+
{\Delta^3\over 2}\ln^2{\Delta_{0}\over\Delta}
\ee
The analysis of the above formulas and the corresponding plot of Fig.3 reveals
a local maximum that resides at $\Delta_0e^{-14/3}$ for $\mu=0$ 
and gradually moves rightward as $\mu$ increases. In contrast,
the global minimum located at $\Delta_0$ first comes down slightly and then 
rises with increasing $\mu$ but, otherwise, retains its position up to
$\mu_c\approx 0.6\Delta_0$.

Upon approaching $\mu_c$ the free energy
computed with the use of the approximate solution (18) shows a behavior suggestive 
of a first order transition. This observation should be contrasted 
with the conjecture that for $\mu=0$ the corresponding transition
is of infinite order.\cite{appelquist}

The above value of $\mu_c$ differs from that obtained in 
Ref. 4, in which a first order
transition was found at $\mu_c^\prime\approx 1.2\Delta_0$. It should be noted, however, that in
their calculation of the free energy, the authors of Ref. 4
introduced the chemical potential as the lower
limit in the integration over $\xi_p$ rather than $E_p$,
which assignment becomes inaccurate for $\mu\sim\Delta_0$.

Lastly, in Fig.4 we present the free energy for $\alpha=1/2$ and $\beta=1$, which 
choice of the parameter values corresponds to the physically relevant problem of
the Coulomb interacting electrons in an infinite
stack of graphene layers \cite{semenoff,dvk}. 
The behavior exhibited by Fig.4 is still indicative of a first order transition 
at $\mu_c\approx 0.7\Delta_0$.

In fact, a more extensive analysis shows that
the above pattern appears to be common in the entire range of parameters $\beta/3\leq\alpha\leq\beta$.
The maximum gap $\Delta_0$ (Fig.5) and the critical coupling $g_c$ (Fig.6),
plotted as functions of the difference $\beta-\alpha$ agree
well with the analytical results given by Eqns.13 and 14, respectively.
The dependence of the critical value of the chemical potential 
$\mu_c$ (Fig.7) upon that parameter turns out to be somewhat 
less pronounced, though.

Before concluding, we comment on our assumption of a spatially uniform order parameter (that allows one
to focus solely on the dependence of $\Delta$ on the 
relative momentum $\bf p$ of particle-hole pairs). Relaxing this simplifying assumption
enables one to extend the class of solutions and search for possible 
Fulde-Ferrel-Larkin-Ovchinnikov (FFLO) states where the order parameter demonstrates periodic
spatial variations. Although we choose to 
postpone a further analysis of this more general case until
future work, our preliminary results suggest that the FFLO-like states may only provide a viable
alternative to the uniform ground state close to the critical value of the chemical potential, 
thus opening the possibility of a sequence (at least, two) transitions   
that might occur in the vicinity of the aforementioned $\mu_c$.

As far as the possible applications of the above results are concerned,
we believe that they may prove helpful for resolving
a number of outstanding issues pertinent to the physics of the nodal strongly correlated 
electron systems.

For one, the recent advent of spintronics has brought about an upsurge of interest in
the magnetic properties of nonmetallic compounds, including pyrolytic graphite
and other degenerate semimetals which can potentially harbor a latent excitonic instability
responsible for the emergence of a (possibly weak, yet robust) ferromagnetism, 
in accordance with the scenario
put forward in Ref. 7.
Although the other recent suspects for the role of excitonic ferromagnet, hexaborides,\cite{rice}
have been largely disqualified on the basis of the experimentally established extrinsic origin
of the observed ferromagnetic response,\cite{bennett} the reports of a weak ferromagnetism in	
some graphitic samples \cite{graphite} has not yet been properly explained.

Whether the magnetism of graphite is associated with a genuine many-body phenomenon
such as the bulk Peierls (charge and/or spin) instability in gapless semimetals
or it has a more mundane nature related to the presence 
of (non)magnetic bulk and/or surface defects,
there remains a compelling need for a comprehensive analysis of
novel, non-BCS-like, pairing scenarios
in nonmetallic systems with long-range interactions.
In particular, it is important to ascertain the extent to which 
the customary practice (see Refs. 11 and references therein)
of replacing the unscreened Coulomb interaction
with the Hubbard-like on-site repulsion (the only motivation for which
seems to be a greater suitability of the latter for numerical simulations)
would allow one to understand the physics of nonmetallic systems.

Also, the idea of an incipient phase transition
reminiscent of the phenomenon of chiral symmetry breaking
has recently gained an additional popularity after a number of (formally
identical but physically distinct) adaptations of the relativistic
QED$_3$ theory were discussed in such contexts
as the antiferromagnet-to-spin liquid and pseudogap-to-charge/spin
density wave transitions in underdoped cuprates.\cite{qed}

Other examples, including a conjectured second pairing transition
in $d$-wave superconductors \cite{yukawa} or $f$-pairing in dichalkogenides,\cite{antonio} have been discussed in the context of
the (pseudo)relativistic 3D Higgs-Yukawa model.
Despite the differences in their underlying physics, all of the aforementioned
problems appear to be amenable to a unified description
in the framework of Eq.(4), with the parameter values
chosen as $\alpha_{\text{QED}}=\beta_{\text{QED}}=1$ or
$\alpha_{\text{HY}}=0$, $\beta_{\text{HY}}=1$. 
Yet another formally
related pairing problem is that of color superconductivity in QCD$_4$
where the integral kernel in the analog of Eq.(4) behaves as $U(p-q)\propto\ln|p-q|$.

The solutions similar to Eq.(15) were also obtained in the earlier studies
of pairing between incoherent quasiparticle excitations mediated by spin fluctuations
in $d$- and $p$-wave superconductors.\cite{chubukov1} In that context, the energy
(albeit not momentum) dependent gap function was found to satisfy a counterpart
of Eq.(4) with $\alpha=\beta=1/2$. 

In summary, we carried out the analysis of the excitonic pairing instability
in nodal fermion systems, characterized by a
power-law DOS and long-range pairwise interaction potentials.
We found non-BCS solutions of the corresponding gap equation at various 
finite values of the chemical potential.
We also computed the free energy and studied its dependence on the chemical potential,
thus gaining further insight into the nature of the transition in question.
We expect these results to be of interest for a broad range of physical problems
involving nodal fermions.

In particular, we observed that the systems with
shorter-ranged interactions (lower values of $\alpha$) and/or stronger DOS suppression
(higher values of $\beta$) tend to manifest a stronger propensity towards excitonic pairing.
The excitonic order emerges at sufficiently strong repulsive 
couplings and survives up to a finite critical doping by excess
carriers. The observed differencies in the behavior at zero and finite
$\alpha$ should serve as a warning against replacing genuinely
long-ranged (e.g., Coulomb) interactions with short-ranged ones, largely
for the sake of a greater ease of the calculations. 

This research was supported by NSF under Grant DMR-0349881 and, in part, by ARO under
Contract DAAD19-02-1-0049. One of the authors (DVK) acknowledges valuable communications
with A. Chubukov, Y. Kopelevich, and V. Gusynin as well as the hospitality at the
Aspen Center for Physics where some of this work was carried out.

%have to change way cite sources
%using BibTeX:
%\bibliography{exnflref}
%\begin{references}

%or, rather, manually - see eg. http://www-h.eng.cam.ac.uk/help/tpl/textprocessing/teTeX/latex/latex2e-html/ltx-73.html

%these three should probably be omitted:
%\end{references}
%\end{multicols}

%See section 9: want to avoid centering arguments
%Use this if want the figure to fit within the column by default:
%\begin{figure}%there is no need to do explicit centering - see http://authors.aps.org/revtex4/template.aps
%\begin{center} 
%\includegraphics[height=0.4\textwidth,angle=0]{Fig_1.eps} \vspace{0.3cm}
%\includegraphics{Fig_1} % Note took out .eps - cannot tell if it improves anything
%\caption {\label{Fig1}Free energy for $\alpha=\beta=0$ 
%as a function of chemical potential $\mu$ (in units of $\Delta_0$):
%$0$ (black), $0.5$ (green), $1/{\sqrt 2}$ (red), and $1$ (blue).}
%\caption {Free energy for $\alpha=\beta=0$ 
%as a function of chemical potential $\mu$ (in units of $\Delta_0$):
%$0$ (black), $0.5$ (green), $1/{\sqrt 2}$ (red), and $1$ (blue).}
%\end{center}
%\end{figure}

%compare that with (much cleaner)
\begin{figure*}
% Using {figure*} instead of {figure} stretches the text across the page instead of confining it to a column
%	There is no need to do explicit centering
	\includegraphics[height=0.4\textwidth,angle=0]{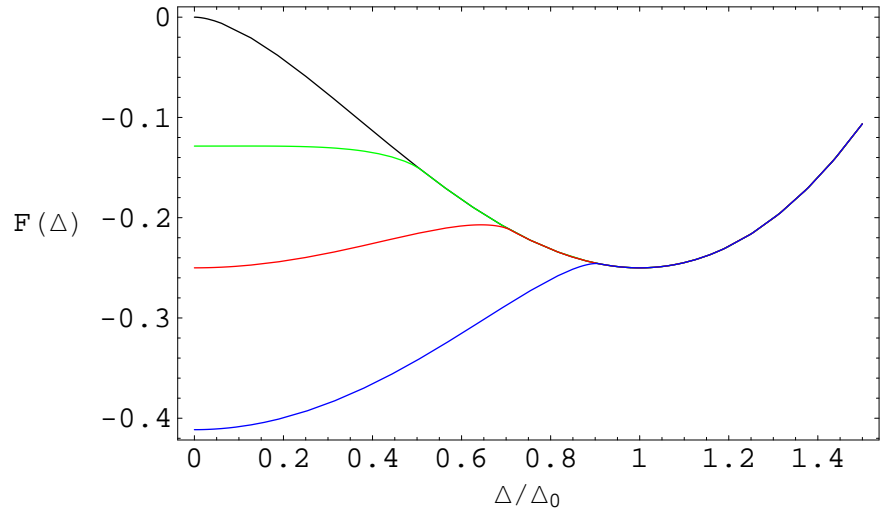}
	\vspace{0.3cm}
	\caption {\label{Fig1}Free energy for $\alpha=\beta=0$ 
as a function of chemical potential $\mu$ (in units of $\Delta_0$):
$0$ (black), $0.5$ (green), $1/{\sqrt 2}$ (red), and $1$ (blue).}
\end{figure*}

%\begin{figure} %allegedly, there is no need to do explicit centering - see http://authors.aps.org/revtex4/template.aps
%\begin{center}
%\includegraphics[height=0.4\textwidth,angle=0]{Fig_2} \vspace{0.3cm}
%\includegraphics{Fig_2.eps}
%\caption {\label{Fig2}Free energy for $\alpha=0$, $\beta=1$ 
%as a function of chemical potential $\mu$ (in units of $\Delta_0$):
%$0$ (black), $0.9$ (green), $1$ (red), and $1.1$ (blue).}
%\end{center}
%\end{figure}

\begin{figure*} 
	\includegraphics[height=0.4\textwidth,angle=0]{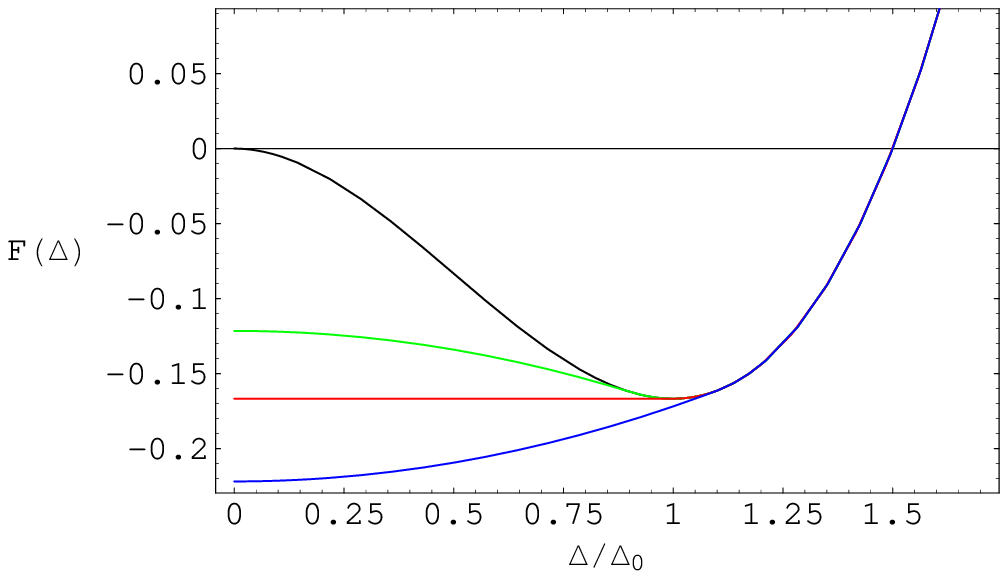} \vspace{0.3cm}
	\caption {\label{Fig2}Free energy for $\alpha=0$, $\beta=1$ 
	as a function of chemical potential $\mu$ (in units of $\Delta_0$):
	$0$ (black), $0.9$ (green), $1$ (red), and $1.1$ (blue).}
\end{figure*}
 
%\begin{figure}
%\begin{center}
%\includegraphics[height=0.4\textwidth,angle=0]{Fig_3.eps} \vspace{0.3cm}
%\caption {Free energy for $\alpha=\beta=1$ 
%as a function of chemical potential $\mu$ (in units of $\Delta_0$):
%$0$ (black), $0.4$ (green), $0.6$ (red), and $0.8$ (blue).}
%\end{center}
%\end{figure}

\begin{figure*}
\includegraphics[height=0.4\textwidth,angle=0]{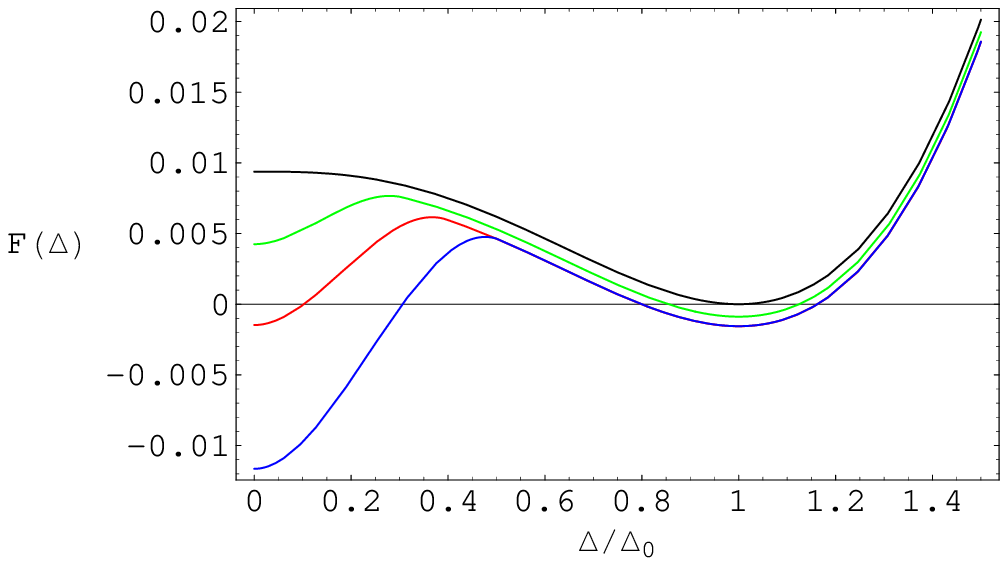} \vspace{0.3cm}
\caption {Free energy for $\alpha=\beta=1$ 
as a function of chemical potential $\mu$ (in units of $\Delta_0$):
$0$ (black), $0.4$ (green), $0.6$ (red), and $0.8$ (blue).}
\end{figure*}

%\begin{figure}
%\begin{center}
%\includegraphics[height=0.4\textwidth,angle=0]{Fig_4.eps} \vspace{0.3cm}
%\caption {Free energy for $\alpha=1/2$, $\beta=1$ 
%as a function of chemical potential $\mu$ (in units of $\Delta_0$):
%$0$ (black), $0.5$ (green), $0.7$ (red), and $0.8$ (blue).}
%\end{center}
%\end{figure}

\begin{figure*}
\includegraphics[height=0.4\textwidth,angle=0]{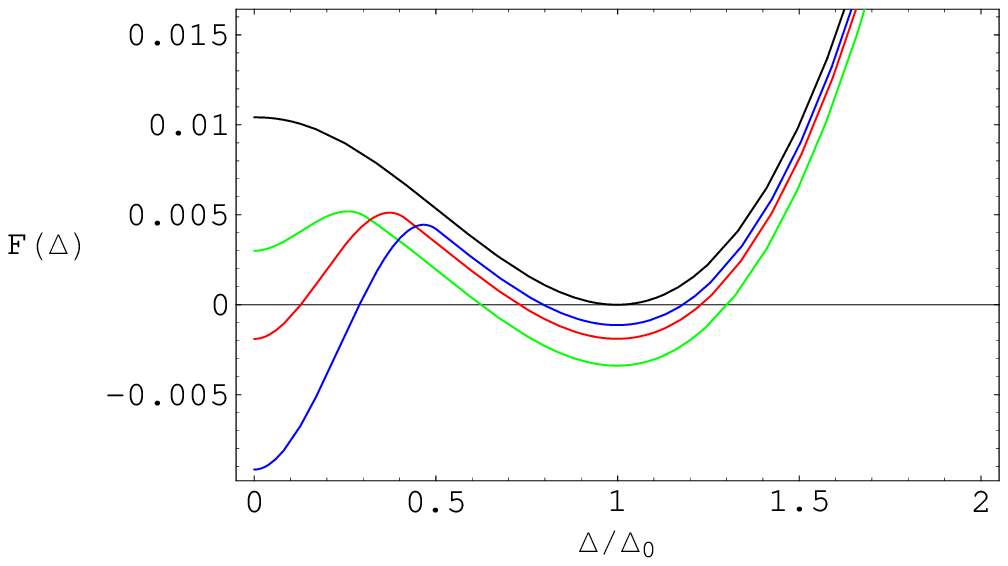} \vspace{0.3cm}
\caption {Free energy for $\alpha=1/2$, $\beta=1$ 
as a function of chemical potential $\mu$ (in units of $\Delta_0$):
$0$ (black), $0.5$ (green), $0.7$ (red), and $0.8$ (blue).}
\end{figure*}

%\begin{figure}
%\begin{center}
%\includegraphics[height=0.4\textwidth,angle=0]{Fig_5.eps} \vspace{0.3cm}
%\includegraphics{Fig_5.eps}
%\caption {Exponent $\nu$ in the maximum gap $\Delta_0\propto ({\tilde g}-{\tilde g}_c)^\nu$
%as a function of $\beta-\alpha$.}
%\end{center}
%\end{figure}

\begin{figure*}
\includegraphics[height=0.4\textwidth,angle=0]{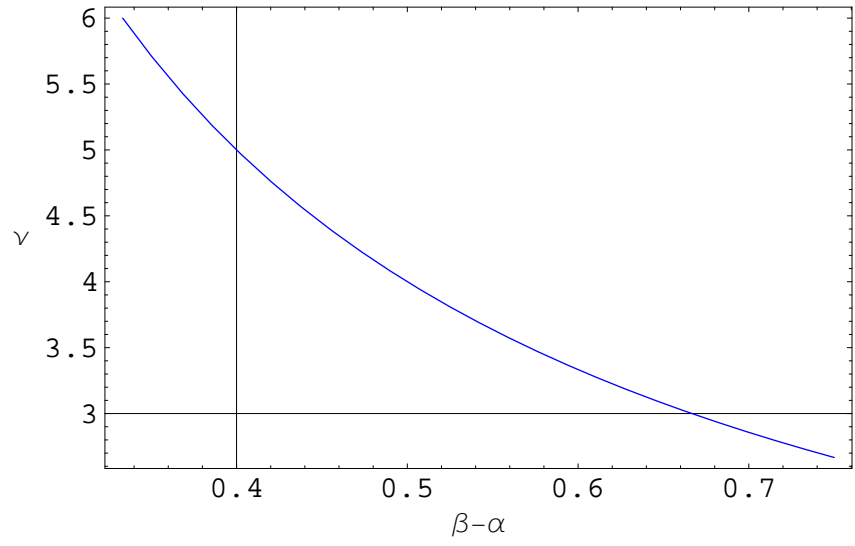} \vspace{0.3cm}
\caption {Exponent $\nu$ in the maximum gap $\Delta_0\propto ({\tilde g}-{\tilde g}_c)^\nu$
as a function of $\beta-\alpha$.}
\end{figure*}

%\begin{figure}
%\begin{center}
%\includegraphics[height=0.4\textwidth,angle=0]{Fig_6.eps} \vspace{0.3cm}
%\includegraphics{Fig_6.eps}
%\caption {Critical coupling ${\tilde g}_c$ as a function of $\beta-\alpha$.}
%\end{center}
%\end{figure}

\begin{figure*}
\includegraphics[height=0.4\textwidth,angle=0]{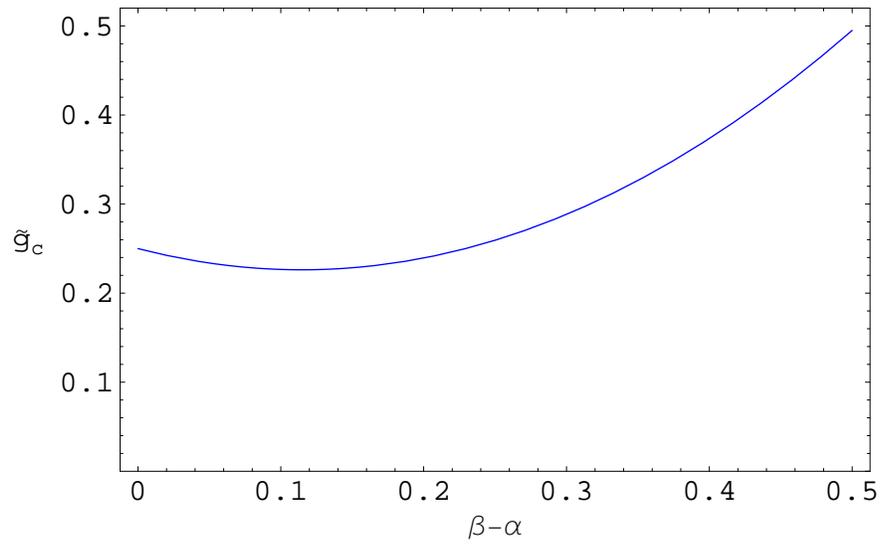} \vspace{0.3cm}
\caption {Critical coupling ${\tilde g}_c$ as a function of $\beta-\alpha$.}
\end{figure*}

%\begin{figure}
%\begin{center}
%\includegraphics[height=0.4\textwidth,angle=0]{Fig_7.eps} \vspace{0.3cm}
%\includegraphics{Fig_7.eps}
%\caption {Critical value of the chemical potential $\mu_c$ (in units of $\Delta_0$)
%as a function of $\beta-\alpha$.}
%\end{center}
%\end{figure}

\begin{figure*}
\includegraphics[height=0.4\textwidth,angle=0]{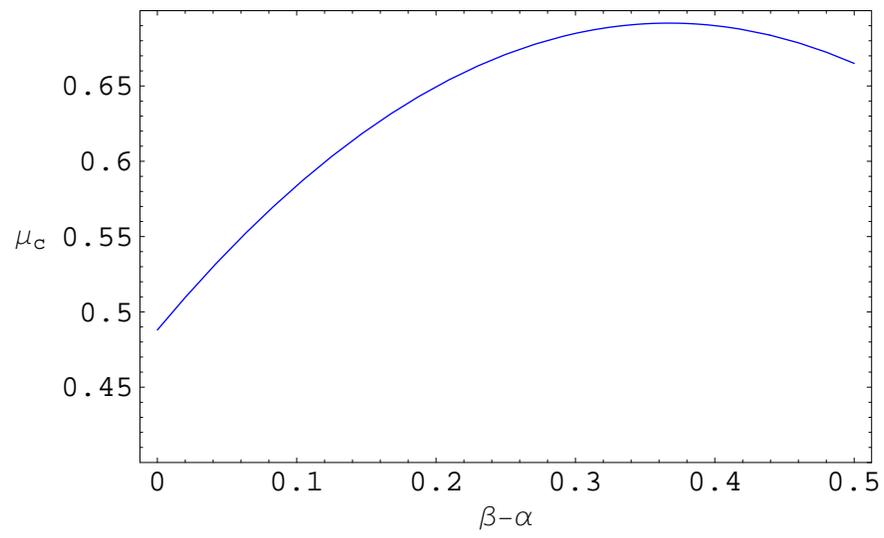} \vspace{0.3cm}
\caption {Critical value of the chemical potential $\mu_c$ (in units of $\Delta_0$)
as a function of $\beta-\alpha$.}
\end{figure*}

\end{document}